\documentclass{article}
\usepackage{graphicx}
\usepackage{amsmath}
\usepackage{amssymb}
\usepackage{amsthm}
\usepackage{bbm}

\usepackage{caption}
\usepackage{subcaption}

\usepackage[maxbibnames=20,sorting=none]{biblatex}
\usepackage{authblk}

\title{\Large On the efficiency of a general attack against the MOBS cryptosystem}
\author{\normalsize Christopher Battarbee, Delaram Kahrobaei, Dylan Tailor and Siamak F.\ Shahandashti}
\affil{Department of Computer Science, University of York, UK \\
\texttt{\{cb203, dk928, siamak.shahandashti\} @york.ac.uk}}

\begin{document}

\maketitle

\begin{abstract}
    All instances of the semidirect key exchange protocol, a generalisation of the famous Diffie-Hellman key exchange protocol, satisfy the so-called ``telescoping equality''; in some cases, this equality has been used to construct an attack. In this report we present computational evidence suggesting that an instance of the scheme called `MOBS (Matrices Over Bitstrings)' is an example of a scheme where the telescoping equality has too many solutions to be a practically viable means to conduct an attack.
\end{abstract}

\section{Introduction}
Since the advent of Shor's algorithm, it has been desirable to study alternatives to the Diffie-Hellman key exchange~\cite{diffie1976new}. One approach to this problem appeals to a more complex group structure: recall that for (semi)groups $G,H$ and a homomorphism $\theta:H\to Aut(G)$, the semidirect product of $G$ by $H$ with respect to $\theta$, $G\rtimes_{\theta}H$, is the set of ordered pairs $G\times H$ equipped with multiplication 
\[(g,h)(g',h')=(\theta(h')(g)g',hh')\]
If $H=Aut(G)$ and $\theta$ is the identity automorphism, we get the holomorph $G\ltimes Aut(G)$. In this case, multiplication has the form 
\[(g,\phi)(g',\phi')=(\phi'(g)g', \phi\circ\phi')\]

The semidirect product can be used to generalise the Diffie-Hellman key exchange \cite{habeeb2013public} via a general protocol sometimes known as the ``non-commutative shift''. Originally, the semigroup of $3\times 3$ matrices over the group ring $\mathbb{Z}_7[A_5]$ is proposed as the platform; however, this turned out to be vulnerable to the type of attack (the so-called ``dimension attack'') by linear algebra described in \cite{myasnikov2015linear},\cite{roman2015linear}\footnote{Both of the terms \textit{non-commutative shift} and \textit{dimension attack} are coined by the authors of \cite{myasnikov2015linear}}. Other platforms used include tropical algebras \cite{grigoriev2014tropical} and free nilpotent $p$-groups \cite{kahrobaei2016using}. The former is shown to be insecure in \cite{isaac2021closer}, \cite{kotov2018analysis}.

In the MOBS protocol \cite{rahmanmobs}, one takes the platform as matrices over the semiring formed by bitstrings under Boolean operations. The hope is that the lack of additive structure in the semiring removes some of the linearity which has left other schemes of this type vulnerable \cite{myasnikov2015linear}, \cite{roman2015linear}. In \cite{brown2021cryptanalysis}, however, one exploits an equation known as the ``telescoping equality''; an approach expanded upon in \cite{battarbee2021cryptanalysis}. The equation allows recovery of a quantity that in some sense encodes information about the private exponents used in key exchange by solving a linear equations. In fact, the telescoping equality is inherent to all schemes of this type; however, the solution is not necessarily unique in the general case. In this report we present computational evidence that the number of solutions to this equation in the case of MOBS is sufficiently large to render a telescoping equality-type attack impractical with the suggested parameters. Indeed, the level of security is considerably better than claimed in the proposal of the scheme.

\section{MOBS}
In the following let $A_i\prod_{i=0}^n$ denote the product $A_mA_{m-1}...A_0$.

The set of $k$-bit strings under the Boolean OR and AND operations forms a semiring; the set of square matrices over this semiring therefore has a notion of multiplication. Call the semigroup of $n\times n$ matrices under this multiplication $S$. Note that any permutation of a $k$-bit string can be extended to a a function on $S$, simply by applying the permutation to each entry of an element of $S$. In fact, doing so yields an automorphism of $S$. Let $M\in S$ and $h$ such an automorphism, and suppose $(M, h)$ is a public holomorph element. Alice and Bob can arrive at a shared secret key as follows:

\begin{enumerate}
    \item Alice picks random $x\in\mathbb{N}$ and calculates $(M,h)^x=(A,h^x)$ and sends $A$ to Bob.
    \item Bob similarly calculates a value $B$ corresponding to random $y\in\mathbb{N}$, and sends it to Alice.
    \item Alice calculates $(B,*)(A,h^x)=(h^x(B)A,**)$ and arrives at her key $K_A=h^x(B)A$. She does not actually calculate the product explicitly since she does not know the value of $*$; however, it is not required to calculate the first component of the product.
    \item Bob similarly calculates his key as $K_B=h^y(A)B$.
\end{enumerate}

Since $A=h^i(M)\prod_{i=0}^{x-1}$, $B=h^i(M)\prod_{i=0}^{y-1}$, we have
    \begin{align*}
        h^x(B)A &= h^x\left(h^i(M)\prod_{i=0}^{y-1}\right)A \\
        &= \left(h^i(M)\prod_{i=x}^{x+y-1}\right)\left(h^i(M)\prod_{i=0}^{x-1}\right) \\
        &= \left(h^i(M)\prod_{i=y}^{x+y-1}\right)\left(h^i(M)\prod_{i=0}^{y-1}\right) \\
        &= h^y(A)B
    \end{align*}
and therefore $K_A=K_B$. We therefore have a secret shared key, denoted $K$, such that $K_A=K=K_B$.    

\subsection{Parameters and Security}
The designers of MOBS suggest parameters $n=3, k=381$. We also require a high-order permutation; to do this we use a product of prime cycles, the order of which will be the product of the primes. $381$ is therefore used as it is the sum of the first $16$ primes.

Infeasibility of key recovery can be reduced to a cousin of the Compuational Diffie-Hellman assumption, which does not appear to be reducible to a discrete-log type problem. We do not give the details as the attack we are interested in is unrelated to solving this security problem.

\section{The Telescoping Equality}
In the following suppose one round of key exchange has been observed by an eavesdropper, and $A$ is fixed corresponding to some fixed $x\in\mathbb{N}$. Using a similar splitting of products as above we can show that 
\[h(A)M=h^x(M)A\]
We will refer to this fact as the telescoping equation or telescoping equality. The idea is that the data $h(A), M, A$ are all available to the eavesdropper; the value $h^x(M)$ apparently encodes some information about the private exponent $x$. In order for this to be useful we need to answer two questions:
\begin{enumerate}
    \item Can we recover $h^x(M)$?
    \item Given $h^x(M)$, can we recover full or partial information about the shared secret key $K$?
\end{enumerate}
The answer to the latter question is yes; we can simply compare $h^x(A)$ and $A$ to recover the permutation $h^x$ by inspection, then calculate $h^x(B)A$, which is the key. 

However, consider the equation $h(A)M=YA$. Certainly $Y=h^x(M)$ is admissible; however, since we are in a semigroup, this is not necessarily a unique solution. The purpose of this investigation is to determine how many solutions there are, thereby judging the feasibility of an attack by the telescoping equality. We also investigate the relationship between the number of solutions to a particular instance of the telescoping equality corresponding to semigroup element $A$, and the size of the left principal semigroup ideal generated by $A$; that is, the size of the set $\{YA: Y\in S\}$.

\section{Experiment Design}
The matrix semigroup is just the direct sum of $k$ semigroups of matrices whose entries are single bits (so-called ``Boolean matrices''). This means that we can decompose the telescoping equality into $k$ single-bit matrix equations, and `reassembling' any combination of the $k$ single-bit solutions will give a solution to the telescoping equality. To find solutions to the single-bit equations we simply try all possible single-bit matrices, of which there are $2^{n^2}$. 

Since Python already allows AND/OR operations on its Boolean values `True' and `False' we represent the matrices as nested lists of `True', `False' values; that is, a random matrix is generated like this:
\begin{verbatim}
    def randstring(k):
        string = []
        for i in range(k):
            m = randint(0,1) 
            if m == 0:
               string.append(False)
            if m == 1:
               string.append(True)
    return string  

    def rand_bool_mat(n,k): 
        matrix = []
        for i in range(n):
            matrix.append([])
            for j in range(n):
                matrix[i].append(randstring(k)) 
    return matrix       
\end{verbatim}

We carry out three experiments, each time counting the logarithm of the number of admissible values in the telescoping equality:

\begin{itemize}
    \item Fix a public matrix $M$ and vary the private exponent $x$
    \item Fix the private exponent $x$ and vary the public matrix $M$
    \item Fix the private exponent $x$ and vary the public matrix $M$, each time counting the size of the left principal ideal generated by $A$
\end{itemize}

Since we will need a list of all single-bit matrices to iterate through, we generate this outside the loop for less expensive computation. The function \textbf{all\_matrices} obtains all length $n^2$ bitstrings from the binary representation of the integers from $0$ to $2^{n^2}-1$, then `folds' them into single-bit matrices.

\begin{verbatim}
    global mats

    mats = all_matrices(3)
\end{verbatim}

In accordance with our general strategy we also need a function to count the number of solutions to single-bit matrix equations. We use the following (where \textbf{prod\_bool\_mat} is just used for the matrix multiplication):

\begin{verbatim}
def count_singlebit_solutions(a, b): 
    count = 0
    for x in mats:
        if a == prod_bool_mat(x, b):
            count = count + 1
    return count
\end{verbatim}

We output the total number of elements $y$ in the preset global iterable \textbf{mats} such that $a = yb$, where $a, b$ are the input matrices. Because any reassembly of single-bit solutions gives a solution to the full equation, the number of solutions to the full equation is the product of the number of solutions to each single-bit equation:

\begin{verbatim}
def count_solutions(a, b):
    ct = 1
    for i in range(len(a[0][0])):
        ct = ct * count_singlebit_solutions(pull(i, a), pull(i, b))
    return ct      
\end{verbatim}

The function \textbf{pull} simply returns the single-bit matrix formed by the $i$-th component of each bitstring entry. We are now ready to define the first of our experiments:
\begin{verbatim}
def count_telescope_solutions_1(M,h): 
    x = randint(2**n, 2**m)
    a = generate_A(M,h,x)
    b = prod_bool_mat(h(a), M)
    return log(count_solutions(a, b))
\end{verbatim}

Since the matrix and matrix permutation are to be fixed and the exponent varied, the matrix and permutation are defined outside of the function. The range of values the exponent can be randomly selected from is defined within the function by the parameters $n, m$.

For the second experiment:
\begin{verbatim}
def count_telescope_solutions_2(h, x, n, k): 
    M = rand_bool_mat(n, k)
    a = generate_A(M,h,x)
    b = prod_bool_mat(h(a), M)
    return log(count_solutions(a, b))
\end{verbatim}

This time we need to input the fixed exponent, and the parameters $n, k$ from which random matrices are generated. We output the logarithm for ease of data visualisation. 

Finally, we need a way to count the size of the left principal ideal generated by $A$; that is, the size of the set $\{YA : Y\in S\}$. We can again exploit the fact that the matrix semigroup is a direct sum of single-bit matrix semigroups: we count the size of the ideal generated by each single-bit matrix, then multiply these numbers. To count the single-bit solutions:

\begin{verbatim}
def count_singlebit_orbit(Y):
    orbit = []
    for x in mats:
        if not check_membership(prod_bool_mat(x, Y), orbit):
           orbit.append(prod_bool_mat(x, Y))
    return len(orbit)    
\end{verbatim}

We can then calculate the size of the full ideal:

\begin{verbatim}
def count_orbit(Y):
    n = len(M)
    k = len(M[0][0])
    orbit_count = 1
    for i in range(k):
        orbit_count = orbit_count * count_singlebit_orbit(pull(i, Y))
    return orbit_count
\end{verbatim}

Keeping the exponent fixed and varying the public matrix, the final experiment is assembled as follows.

\begin{verbatim}
def count_telescoping_solutions_orbit(m_p, m_s, b_l, exp):
    M = rand_bool_mat(matrix_size, bitstring_length)
    a = generate_A(M, matrix_permutation, exponent)
    b = prod_bool_mat(matrix_permutation(a), M)
    return (count_orbit(a), count_solutions(b, a))
\end{verbatim}

\section{Results}
The experiments ran on $3\times 3$ matrices, with three different values of $k$. Each value of $k$ is the sum of the first few primes. The results of the trials suggest that: 
\begin{itemize}
    \item Each matrix $M$ corresponds to a fixed number of solutions to the telescoping equality, regardless of exponent
    \item With the parameters suggested, there are sufficiently many solutions to the telescoping equation for any matrix to make an attack via the telescoping equality infeasible
    \item There is negative correlation between the size of the ideal generated by a particular exchange value $A$ and the number of solutions to the corresponding telescoping equality
\end{itemize}

\subsection{Independence from Exponent}
When the matrix and permutation are fixed but the exponent in the calculation of the exchange value $A$ is varied, over several thousand trials we did not encounter a case where the number of solutions changed. This suggests that the number of solutions to the telescoping equality is independent of the exponent, although we do not have an explanation for this behaviour.

For our purposes, assuming that independence from exponent does indeed hold, we conclude that we can run the remaining experiments on small (arbitrarily we decide on 100) exponents for less expensive computation; that is, we do not need to use the large parameters suggested by the authors of MOBS\footnote{In fact, we find that the number of solutions is dependent on exponent for very small values of exponent, but stabilise after a while to independence of exponent; we therefore choose a fixed exponent to balance low computational cost with surpassing this boundary}. However, we note a curious detail: with a small tweak to the experiment \textbf{count\_telescoping\_solutions\_1} we can output a list of the number of solutions to the single-bit equations, rather than their product (the original output). Changing the exponent seems to permute the output list; for example, a $3\times 3$ matrix with $5$-bit entries chosen at random gives the following: the table gives the exponent used for calculation in its first column, then the number of solutions to each single-bit matrix equation in order. 

\begin{center}
    \begin{tabular}{c|c|c|c|c|c}
        \textbf{28} & 49 & 72 & 72 & 72 & 216 \\
        \textbf{67} & 49 & 72 & 72 & 72 & 216 \\
        \textbf{89} & 72 & 72 & 216 & 72 & 49 \\
        \textbf{96} & 216 & 72 & 49 & 72 & 72 \\
        \textbf{43} & 49 & 72 & 72 & 72 & 216 \\
        \textbf{98} & 72 & 72 & 215 & 72 & 49 \\
        \textbf{36} & 216 & 72 & 49 & 72 & 72 \\
        \textbf{84} & 216 & 72 & 49 & 72 & 72 \\
        \textbf{64} & 49 & 72 & 72 & 72 & 216 \\
        \textbf{63} & 216 & 72 & 49 & 72 & 72 
    \end{tabular}
\end{center}

\subsection{Number of Solutions}
\begin{figure}
     \centering
     \begin{subfigure}[b]{0.3\textwidth}
         \centering
         \includegraphics[width=\textwidth]{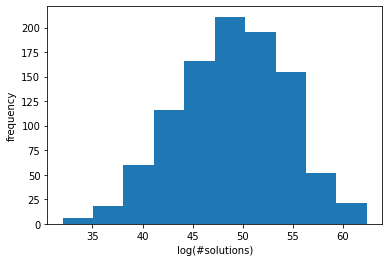}
         \caption{$k=10$}
         \label{fig:y equals x}
     \end{subfigure}
     \hfill
     \begin{subfigure}[b]{0.3\textwidth}
         \centering
         \includegraphics[width=\textwidth]{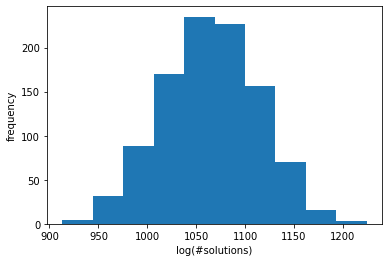}
         \caption{$k=197$}
         \label{fig:three sin x}
     \end{subfigure}
     \hfill
     \begin{subfigure}[b]{0.3\textwidth}
         \centering
         \includegraphics[width=\textwidth]{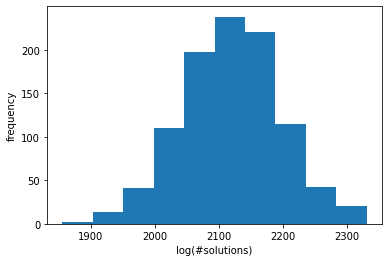}
         \caption{$k=381$}
         \label{fig:five over x}
     \end{subfigure}
        \caption{Number of Solutions}
        \label{fig:solutions}
\end{figure}

Figure \ref{fig:solutions} shows a histogram of the logarithm of the number of solutions to the telescoping equality when the exponent and permutation are fixed and the public matrix is varied, conducted over a thousand trials. The key takeaway is that in all trials there are far too many solutions to make recovering the correct one a viable strategy; for the suggested parameter $k=381$ even the smallest number of solutions is in the range of $2^{1900}$, and this number of solutions did not occur frequently. The histograms also seem to suggest that the number of solutions are roughly normally distributed within their range.

\subsection{Number of Solutions vs. Ideal Size}
\begin{figure}
     \centering
     \begin{subfigure}[b]{0.3\textwidth}
         \centering
         \includegraphics[width=\textwidth]{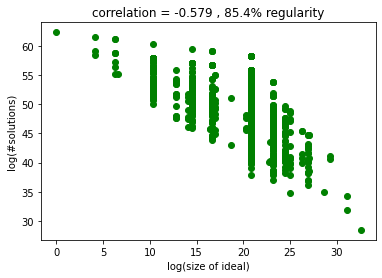}
         \caption{$k=10$}
         \label{fig:y equals x}
     \end{subfigure}
     \hfill
     \begin{subfigure}[b]{0.3\textwidth}
         \centering
         \includegraphics[width=\textwidth]{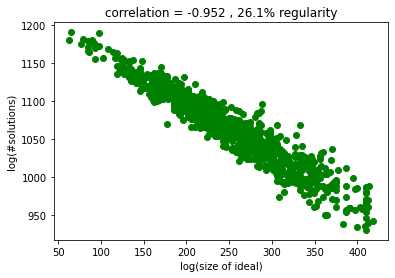}
         \caption{$k=197$}
         \label{fig:three sin x}
     \end{subfigure}
     \hfill
     \begin{subfigure}[b]{0.3\textwidth}
         \centering
         \includegraphics[width=\textwidth]{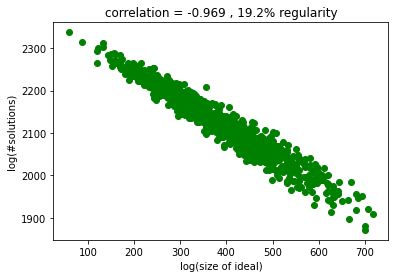}
         \caption{$k=381$}
         \label{fig:five over x}
     \end{subfigure}
        \caption{Orbits vs. Number of Solutions}
        \label{fig:orbits v. solutions}
\end{figure}

Figure \ref{fig:orbits v. solutions} shows the logarithm of the number of solutions against the logarithm of the size of the principal ideal generated by the corresponding value of $A$. The tests were conducted over a thousand trials with $3\times 3$ matrices; for each trial, the matrix permutation and exponent are kept constant. The graphs exhibit reasonably strong negative correlation, which one would naively expect - a larger ideal means that $YM$ lands on the quantity in the telescoping equality less frequently. The vertical lines in graph (a) show that for two matrices whose corresponding exchange value $A$ has the same ideal size, their corresponding telescoping equality does not necessarily have the same number of solutions. Indeed, these vertical lines would be present on the other two graphs at higher resolution. We also point out that experiments into ideal size when exponent was varied and other parameters fixed yield similar results to those when number of solutions is counted.

At the top of each graph two data points are noted: first, Spearman's correlation coefficient, second, the percentage of data points achieving regularity. We say a bitstring matrix $M$ is \textbf{regular} if the number of $Y$ satisfying $h(A)M=YA$ is the same as the number of $Y$ satisfying $Yh(A)M=A$. We note that as the bitstring length increases we get better correlation and worse regularity rates.

\section{Conclusions}
\subsection{Security Implications}
We know of no better way of identifying which of the quantities satisfying the telescoping equality is the correct value to conduct an attack than simply guessing. The large number of solutions in our results suggest that the probability of choosing the correct value is vanishingly small. Moreover, the technique of decomposition and reassembly reduces the number of matrix multiplications from $2^{kn^2}$ to $k2^{n^2}$. As far as we know, then, the methods described above are the optimal method of conducting an attack by the telescoping equality. We conclude that the basic attack by the telescoping equality will not work against this scheme.

However, given that there appears to be negative correlation between ideal size and number of solutions, one should be careful to choose a public matrix such that the corresponding exchange values $A$ generate small ideals. This is more pertinent should one wish to use smaller parameter sizes.

\subsection{Relationship to Monico Attack}
In \cite{cryptoeprint:2021:1114} a polynomial-time attack on the MOBS protocol related to our discussion is given. The strategy is to find an integer $a$ such that $h^a(M)A=h(A)M$; one certainly exists by the telescoping equality. It turns out that such an $a$ will satisfy $h^a(B)A=h^y(A)B=K$. One can effectively find such an $a$ by combining the fact that the permutation $h$ is made up of disjoint prime-order cycles, and that the matrix group decomposes as a direct sum in the way we have discussed throughout this report. 

Towards generalising the attack, for any public automorphism $\phi$ Monico points out that one need only find an automorphism $\psi$ such that $\psi$ commutes with $\phi$, and the equality $\psi(g)A=\phi(A)g$ holds. It is not clear that MOBS has any inherent vulnerability to recovery of such a $\psi$; rather, we exploit the structure of $h$ to find an example of such a $\psi$. Should a different automorphism be used, therefore, the methods of \cite{cryptoeprint:2021:1114} do not immediately guarantee an effective method of key recovery. However, as our experimental results suggest that a public pair $(M, h)$ corresponding to an exchange value $A$ with large ideal size is bad for security, one must be careful to balance these considerations.

\section{Acknowledgement} We thank Vladimir Shpilrain for helpful comments and discussions.
\printbibliography

\end{document}